\documentclass{iopart}
\usepackage{iopams} 
\usepackage{slashed}
\usepackage{color}
\usepackage{graphicx}
\newtheorem{theorem}{Theorem}
\newtheorem{definition}{Definition}
\newtheorem{corollary}{Corollary}
\newtheorem{lemma}{Lemma}

\def\Dd{D_A-\mu}
\def\di{\pndolch-\mu}
\def\fid{\varphi^{(\delta)}}
\def\la {\Lambda_{A}}
\def\Sd{S^{(A)}}

\def\Wd{W^{(\delta)}}
\def\Xd{X^{(\delta)}}
\def \pdolch {\slashed{\bp}}
\def \pndolch {\slashed{\bp}_0}
\def\tr{\mathop{\mathrm{tr}}\nolimits} % Spur
\def \bp {\mathbf{p}}
\def \bA {\mathbf{A}}
\def \ri {\mathrm{i}}
\def\he#1{#1^{\frac12}}
\def\mhe#1{#1^{-\frac12}}
\def\xinn{\xi_\nu^{(n)}}
\def \hf {h_{\mathrm{HF}}}
\def \core { C_0^\infty(\rz^2:\cz^2)}
\def \ra {R_A(z)}
\def \rp {R_0(z)}

\newcommand{\cE}{\mathcal{E}}
\newcommand{\cS}{\mathcal{S}}

\newcommand{\gB}{\mathfrak{B}}

\newcommand{\gH}{\mathfrak{H}}

\newcommand{\gS}{\mathfrak{S}}

\newcommand{\rd}{\mathrm{d}}
\def \ri {\mathrm{i}}

\newcommand{\bB}{\mathbf{B}}

\newcommand{\bx}{\mathbf{x}}
\newcommand{\by}{\mathbf{y}}

\newcommand{\cz}{\mathbb{C}} % Komplexe Zahlen
\newcommand{\gz}{\mathbb{Z}} % Ganze Zahlen
\newcommand{\nz}{\mathbb{N}} % Nat"urliche Zahlen
\newcommand{\rz}{\mathbb{R}} % Relle Zahlen
 
\begin{document}

\title[Multiparticle equations for graphene]{Multiparticle 
equations for interacting Dirac fermions in 
magnetically confined graphene quantum dots}

\author{Reinhold Egger$^1$, Alessandro De Martino$^2$,
 Heinz Siedentop$^3$, and Edgardo Stockmeyer$^3$}

\address{
$^1$~Institut f\"ur Theoretische Physik, Heinrich-Heine-Universit\"at, 
Universit\"atsstra{\ss}e 1, D-40225 D\"usseldorf, Germany\\
$^2$~Institut f\"ur Theoretische Physik, 
Universit\"at zu K\"oln, Z\"ulpicher Stra{\ss}e 77, D-50937 K\"oln, Germany\\
$^3$~Mathematisches Institut, 
Ludwigs-Maximilians-Universit\"at M\"unchen, 
Theresienstra{\ss}e 39, D-80333 M\"unchen, Germany}
\eads{ \mailto{egger@thphy.uni-duesseldorf.de},
\mailto{ademarti@thp.uni-koeln.de},
\mailto{h.s@lmu.de}, \mailto{stock@math.lmu.de}}

\begin{abstract}
We study the energy of quasi-particles in graphene within the
Hartree-Fock approximation. The quasi-particles are confined via an
inhomogeneous magnetic field and interact via the Coulomb potential.
We show that the associated functional has a minimizer and
determines the stability conditions for the $N$-particle
problem in such a graphene quantum dot.
\end{abstract}

\submitto{\JPA}
%\maketitle

\section{Introduction}\label{section1}

The electronic properties of graphene -- a two-dimensional monolayer
of graphite made of carbon atoms only -- have recently attracted a lot
of interest \cite{review1,review2,review3}.  For energies close to the
charge neutrality point, noninteracting quasi-particles in graphene
(henceforth called ``electrons'') are well described by the Dirac-Weyl
Hamiltonian of massless relativistic fermions. 
This suggests an easily accessible condensed-matter
realization of relativistic quantum mechanics.  Recent interest has
turned to Coulomb interaction effects, in particular to the case when
$N$ electrons are confined to a finite region (a so-called ``quantum
dot'') in the graphene layer.  Using electrostatic confinement
potentials, such Coulomb-correlated artificial atoms have been studied
in detail for the two-dimensional electron gas in semiconductors
\cite{reimann}, where the non-relativistic Schr\"odinger equation
describes the single-particle sector.  In graphene, however, the Klein
tunnelling phenomenon \cite{review2} of relativistic Dirac fermions
renders the standard electrostatic confinement method experimentally
difficult or even impossible to apply, and usually only resonances
 can be expected \cite{th1,th2,th3,th4,th5}.  As an alternative,
confinement by inhomogeneous orbital magnetic fields has been
suggested \cite{ademarti}, and the electronic structure of two
interacting electrons in such a magnetic quantum dot in graphene was
recently studied using exact diagonalization \cite{hausler}.
Experimentally, up to now, lithographically fabricated quantum dots
were mostly studied \cite{ensslin}, where the boundary is rather
disordered, the confinement potential cannot be tuned, and a detailed
comparison of experimental data to theory is difficult.  On the other
hand, inhomogeneous magnetic fields have been experimentally generated
and studied in semiconductor devices by using suitable
lithographically deposited ferromagnetic layers \cite{heinzel}, and
the generalization to graphene should pose no fundamental obstacle.
Concrete experimentally relevant profiles for quantum dot confinement
by such fields were theoretically studied also in Ref.~\cite{rei}.  We
note that an artificial vector potential giving rise to the same
mathematical model can also be generated by applying mechanical
forces, producing appropriate strain in the sample \cite{review2,review3}.

In this paper, given the widespread interest in understanding and
usefully employing the electronic structure of graphene quantum dots,
we address the definition and the stability of the relativistic
interacting $N$-particle system in such a magnetic graphene dot,
primarily from a mathematically oriented perspective.  To that end, we
analyze in detail the Hartree-Fock functional and show that under
certain conditions,  a minimizer exists.  The
maximum number $N_c$ of particles is computed and shown to depend on
the interaction strength. When $N\leq N_c$ we solve the Hartree-Fock
equations numerically. For $N>N_c$, no minimizer with particle
number $N$ exists. The excess particles are not localized and 
appear, numerically, occupying bulk
Landau states centered far away from the dot region to lower the repulsive
interaction energy. 

It is well known that for relativistic $N$-particle problems,
the presence of interactions implies that one should 
use a projection scheme \cite{Sucher1980,Sucher1984,Sucher1987}, 
whose integrity for graphene dots -- despite the fact that bulk graphene
corresponds to a gapless model -- has been shown in Ref.~\cite{hausler}.
At first sight, an inclusion of the entire Dirac sea along the more
fundamental lines of Hainzl \textit{et al.} (see \cite{hainzl} and the
references therein) might seem desirable. However, this would
complicate matters unduly, since -- after all -- the description of
graphene by a two-dimensional Dirac equation emerges from a
non-relativistic band structure calculation. Moreover, in the presence
of a finite gap separating occupied and empty states -- which is the
case below -- it is not only reasonable to freeze the Dirac sea as
given by the external field (Furry picture), but also to assume that no
electron-positron pairs are created. It should be remarked that this
strategy is not only supported by Ref.~\cite{hausler} but is a standard
procedure in quantum chemistry \cite{ReiherWolf2009}.  Because of
this we will restrict ourselves in this work to the no-pair model, more
precisely to the no-pair model in the Furry picture with given electron
number $N$.

In this model, the electronic states of graphene in the presence of a
magnetic field $\nabla\times\tilde\bA$ are the unit vectors in
$\chi_{(0,\infty)}(v\boldsymbol\sigma\cdot(-\ri\hbar\nabla+\case{e}{c}
\tilde\bA))[L^2(\rz^2:\cz^2)]$, where $\tilde\bA$ is the magnetic
vector potential, $v$ is the Fermi velocity in graphene,
$\boldsymbol\sigma=(\sigma_1,\sigma_2)$ are the first two Pauli
matrices, $-e<0$ the charge of the electron, and $\hbar$ the Planck
constant.  We introduce convenient units by scaling momentum $\bp
\mapsto \bp/(v\hbar)$ and coordinates $\bx \mapsto v\hbar\bx$, which
generates a unitary transform.  $N$ Coulomb-interacting electrons in
graphene are then formally described by the Hamiltonian
\begin{equation} \label{physikhamilton}
  \sum_{n=1}^N \left[ \boldsymbol\sigma_n\cdot(\bp_n+\mathbf{A}(\bx_n)) 
-\varphi(\bx_n) \right] + \sum_{1\leq m<n\leq N}\frac{\alpha}{|\bx_m-\bx_n|}
\end{equation}
projected onto the antisymmetric tensor product of the above space.
Here $\mathbf{A}(\bx):= (ev/c)\tilde\bA(v\hbar\bx)$ and we have 
added an external electrostatic potential $\varphi$.  The interaction
strength  is encoded in the dimensionless fine structure constant
$\alpha=e^2/(v\kappa\hbar)$.
  Physical values in graphene are $0<\alpha< 2$, with 
the precise value depending on the dielectric constant $\kappa$ of the
environment (e.g., the substrate on which graphene is deposited). 
The upper limit for $\alpha$ is approached only
for suspended samples.  It is well known that quasi-particles in 
graphene have additional valley (``$K$ point'') and
 electronic spin degrees of freedom.
For field configurations that are smooth on the scale of graphene's
lattice constant $a_0=0.246$~nm, however, no valley mixing
is expected, and our simpler description with just one $K$ point 
and full spin polarization is sufficient \cite{hausler}.  
This is the situation considered in our paper.  In any case, 
the generalizations necessary to go beyond the single-spin and 
single-valley model (\ref{physikhamilton}) are conceptually unproblematic.
A similar approach is also expected to apply for 
magnetic dots in bilayer graphene, where the linear dispersion
relation of monolayer graphene is modified \cite{review2}.

The outline of the remainder of this paper is as follows.  In 
Sec.~\ref{sec2}, we specify and explain the mathematical model. 
The Hartree-Fock functional is considered in Sec.~\ref{sec3}, and we prove the
existence of a minimizer.  For the benefit of the mathematically
oriented reader, we have included all proofs in detail. 
To make the abstract discussion in Sec.~\ref{sec3} concrete, 
we describe a specific magnetic field profile $\mathbf{A}(\bx)$
in Sec.~\ref{sec4}, modelling a circular magnetic dot of radius $R$ around
the origin.  The confinement is here generated by a constant magnetic
field $B$ for $r>R$ but zero field for $r<R$.  The relativistic interacting
$N$-particle problem for a magnetic dot in graphene
is then studied for this field profile in Sec.~\ref{sec4} by
 numerically solving the Hartree-Fock equations discussed in 
Sec.~\ref{sec3}.  This calculation also yields quantitative 
predictions concerning the stability of the $N$-electron system.
We conclude with an outlook in Sec.~\ref{sec5}. 
Some technical details have been relegated to two Appendices.

\section{Model}
\label{sec2}

To describe the physical situation where electrons are bound to a 
magnetic quantum dot, the magnetic vector potential $\bA=\bA_0 + \bA_1$
is expressed as the sum of a homogeneous magnetic field of
strength $B>0$ perpendicular to the graphene plane,
$\bA_0(\bx)= (B/2) (-x_2,x_1)$, and a perturbation $\bA_1$.  
For total external electromagnetic field $A:=(\varphi,\bA)$, 
using $\pndolch := \boldsymbol\sigma\cdot(\bp+\bA_0)$, we define 
\begin{equation}
D_A := \pdolch - \varphi := \boldsymbol\sigma\cdot(\bp+\bA)-\varphi 
= \pndolch + P 
\end{equation}
with the ``perturbation'' $P:= \boldsymbol\sigma\cdot\bA_1 -\varphi$.  
We define the one-electron Hilbert space with respect to $A$ as
\begin{equation}
\gH_A:=\la^+[L^2(\rz^2:\cz^2)] \;, \quad \la^+:= \chi_{(0,\infty)}(D_A) \; .
\end{equation}
Although more general cases can be treated, we assume for simplicity
that $P$ is a bounded operator which is sufficient for the application
discussed and minimizes the amount of technical arguments
needed. In the same spirit, we require relative
compactness of the perturbation\footnote{For $p\in [1,\infty)$, we set
    $\gS^p(\gH_A)=\{ A\in\mathcal{B}(\gH_A) \big| \tr|A|^p <\infty\}$,
    write $\gS^\infty(\gH_A)$ for the space of compact, and
    $\gB(\gH_A)$ for the bounded operators on $\gH_A$.}
\begin{equation} \label{eq:P}
  P\in \gB(\gH_A)\quad {\rm and}\quad
 \he{|P|}\mhe{|\di|}\in\gS^\infty(\gH_A)\;.
\end{equation}
Physically, this means that the perturbing electromagnetic potential
decays at infinity and has only moderate singularities whose
Fourier coefficients can be controlled by the kinetic energy.
At the same time, $P$ is responsible for creating bound states 
that define the quantum dot.

The energy of an $N$-particle state
$\psi\in\cS(\rz^{2N}:\cz^{2^N})\cap\gH_\bA^{(N)}$\footnote{Here $\cS$
  denotes the space of Schwartz functions, and $\gH_\bA^{(N)}$ is the
  $N$-fold antisymmetric tensor product of the one-electron Hilbert
  space $\gH_\bA$, i.e., the canonical $N$ electron space.}  
is then given by
\begin{equation} \label{energie}
  \cE(\psi) := \left (\psi, \left[\sum_{n=1}^N (D_{A,n}-\mu)
  + \sum_{1\leq m<n\leq N} \frac{\alpha}{ | \bx_m-\bx_n|} \right]\psi\right)\;.
\end{equation}
Throughout this paper, $\mu$ is a positive constant smaller than the
first positive eigenvalue (first bulk Landau level) of $\pndolch$.  Of
course, since $\mu$ is just a constant,  the energy is shifted
merely by $-\mu\tr\gamma$. However, this shift serves an important
technical purpose: it will allow us to replace the minimization under
the constraint $\tr \gamma \leq N$ instead of $\tr \gamma =N$. If the
minimizer $\gamma$ has trace $N$, then the problem with the $N$
electron constraint has been solved. Note, that it may happen that the
trace of the minimizer, $\tr\gamma$, stays always below some $n_c$, even
if $\mu$ is close to the Landau level. Then the smallest such $n_c$ --
call it $N_c$ -- is the maximal number of electrons which can be
captured by the dot, see Sec.~\ref{sec4}.  For $N>N_c$, there are
$N-N_c$ unbound electrons. The case that there is no minimizer with
trace equal to $N$ corresponds exactly to the situation, where those
$N-N_c$ electrons float away to infinity and cannot be bound by the dot.

The form (\ref{energie}) is obviously bounded from below and closable. 
This allows us to define the Hamiltonian $B_A$ as its Friedrichs extension.
The Friedrichs extension is the canonical way to construct a self-adjoint
Hamiltonian out of a symmetric operator which is bounded from below,
see, e.g., Refs.~\cite[Satz~4.15]{weidmann} or \cite[Theorem~X.23]{reedsimon}. 

\section{The Hartree-Fock Functional}\label{sec3}

\subsection{The Hartree-Fock Functional\label{s2.1}}

We are now interested in the Hartree-Fock approximation for the
ground state of the Hamiltonian $B_A$.  The Hartree-Fock 
method is a standard tool to access interaction effects in 
atomic, molecular, or condensed-matter systems 
\cite{ReiherWolf2009,vignale}. The
 Hartree-Fock ground-state energy provides an upper bound for
the true ground-state energy corresponding to Eq.~(\ref{energie}),
and it can be used to assess the stability of the $N$-particle
problem \cite{stability}.
We mention that a different variational approach based on the so-called
M\"uller functional can yield lower bounds for the ground-state
energy \cite{muller,siedentop}, and very useful results can be obtained by
combining both methods. 

With the above choice for $\mu$, let us denote
by $d$ the distance of $\mu$ to the nearest spectral point of $\pndolch$. 
We start our analysis by defining the basic class of operators $(\gamma$) that 
enter the Hartree-Fock functional.  In physical terms, $\gamma$ is
the density operator. 
\begin{definition} \label{def:F} We define the Banach space
  \[
    F:=\{\gamma\in\gB(\gH_A)\big|
  \|\gamma\|_F :=\||\di|^{1/2}\gamma|\di|^{1/2}\|_1<\infty,\
 \gamma=\gamma^*\} 
  \]
\end{definition}
Here, as customary, $\|a\|_1 := \tr\sqrt{a^*a}$ denotes the trace norm of
the operator $a$.  This definition of $F$ is motivated\footnote{For
  later use of the Banach-Alaoglu theorem, we also note that
\[
F_*:= \left\{\delta\in\gB(\gH_A)\big||\di|^{-\frac12}\delta 
|\di|^{-\frac12}\in\gS^\infty(\gH_A)\right\}
\]
is a Banach space for which $F$ is the dual space. The duality is
given naturally as
\[
 \langle\gamma,\delta\rangle
:= \tr \left(|\di|^{\frac12}\gamma |\di|^{\frac12}
|\di|^{-\frac12}\delta |\di|^{-\frac12} \right) = \tr (\gamma\delta) \;.
\]}
by the fact that the state represented by $\gamma$ should have
 finite kinetic energy, finite particle number, and 
real occupation numbers, the latter being the eigenvalues of $\gamma$.  
Note that $\gS^1(\gH_A) \supset F$ since
\begin{eqnarray*}
  \|\gamma\|_F & = & \tr\sqrt{|\di|^{\frac12}\gamma|\pndolch-
\mu|\gamma|\di|^{\frac12}}\\ &\geq& d^{\frac12} \tr 
\sqrt{|\di|^{\frac12}|\gamma||\gamma||\di|^{\frac12}}\\
  &=& d^{\frac12}\tr\sqrt{ |\gamma| |\di| |\gamma|} \geq d\tr |\gamma|\; .
\end{eqnarray*}
For a given element $\delta\in F$, we denote its eigenvalues by
$\lambda_n$ and its eigenspinors by $\xi_n$. The associated integral
kernel $\delta(x,y)$ is\footnote{We use the notation $x=(\bx ,s)$ for an
element of $G:=\rz^2\times\{ 1,\,2\}$ and $\rd x$ for the
product of the Lebesgue measure on $\rz^2$ with the counting measure
in $\{1,\,2\}$.}
\begin{equation} \label{eq:kern}
\delta(x,y):=\sum_n \lambda_n \xi_n(x)\overline{\xi_n(y)}\;.
\end{equation}
Associated with $\delta$ is its one-particle density
\[
\rho_\delta(\bx) := \sum_{s=1}^2\sum_n \lambda_n |\xi_n(x)|^2\;,
\]
its electric potential $\fid(\bx) := \int \rd \by \
\rho_\delta(\by)/|\bx-\by|$, 
and its exchange operator $\Xd$ in terms of the integral kernel
$\Xd (x,y)=   \delta(x,y)| \bx - \by|^{-1}$.  Then
$\Wd=\fid-\Xd$ is the mean field potential, and the Weyl operator 
associated to $\delta$ is 
\begin{equation}\label{D-F-operator}
    \hf^{(\delta)} := \Dd+ \alpha \Wd \;.
\end{equation}
The Coulomb scalar product is defined as
\begin{equation}
  \label{coulomb}
  D(\rho,\sigma):=\frac12 \int_{\rz^2}\rd\bx \int_{\rz^2} \rd\by
  \frac{\overline{\rho(\bx)}\sigma(\by)}{| \bx - \by|}\;,
\end{equation}
and the exchange scalar product for $\gamma,\gamma'\in F$ is
\begin{equation}
  \label{eq:austausch}
  X(\gamma,\gamma') :=\frac12\int\rd x\int\rd y
  {\overline{\gamma(x,y)}\gamma'(x,y)\over|\bx-\by|} \;.
\end{equation}
Indicating quadratic forms of sesquilinear forms by brackets, 
e.g., $X(\gamma, \gamma)=X[\gamma]$, we set
$Q[\gamma]:= D[\rho_\gamma]- X[\gamma]$.

We now discuss inequalities which will allow us to
define the Hartree-Fock energy functional. 
\begin{lemma} \label{Ede-Ungleichung}
We have
\[ 
|\bp+\bA_0| \leq  |\pndolch| + B^{1/2}\leq |\di|+\mu+B^{1/2}\;.
\]
\end{lemma}
{\it Proof.} We have
\[
  (\psi,(\bp+\bA_0)^2 \psi)
  \leq (\psi, ((\bp+\bA_0)^2 +\boldsymbol \sigma \cdot \bB + B)\psi)
  =(\psi, (\pndolch^2 +B)\psi) \;.
\]
Using that the root is operator monotone and $\sqrt{\pdolch_0^2 +B}
 \leq |\pdolch_0| + B^{1/2}$, the first inequality follows.
The second inequality is then clear.  $\hfill \Box$

\begin{lemma}
  \label{sec:definition-problem}
  Assume that $\gamma,|\gamma'|\in F$. Then
  \begin{eqnarray}
    \label{eq:q1}
    |D(\rho_\gamma,\rho_{\gamma'})| &\leq& (2h)^{-1}
     \|\gamma\|_1 \left [ \||\gamma'|\|_F+(\mu+B^{\frac12})\|\gamma'\|_1
\right]\;,\\
    \label{eq:3}
    |X(\gamma,\gamma')|&\leq& D\left(\rho_{|\gamma|},\rho_{|\gamma'|}\right)\;.
  \end{eqnarray}
\end{lemma}
{\it Proof.} Expanding $\gamma$ and $\gamma'$ in eigenfunctions,
see Eq.~(\ref{eq:kern}), we get by the Schwarz inequality
\begin{eqnarray*}
  &&  \left|\int\int \frac{\overline{\gamma(x,y)}\gamma'(x,y)}{|
        \bx-\by|}\rd x\, \rd y\right|  \\
 && = \left|\int \int \sum_\mu
      \lambda_\mu \sum_\nu \lambda'_\nu
      \frac{\overline{\xi_\mu(x)}\xi_\mu(y)\xi'_\nu(x)\overline{\xi'_\nu(y)}}
      {|\bx-\by|}\rd x \rd y \right|\\
   &&  \leq \int \int \frac{\sum_\mu|\lambda_\mu| |\xi_\mu(x)|^2
      \sum_\nu|\lambda'_\nu||\xi'_\nu(y)|^2} {|\bx-\by|}\rd x \rd y \\
 && = \int_{\rz^2} \int_{\rz^2}
    \frac{\rho_{|\gamma|}(\bx)\rho_{|\gamma'|}(\by)}{| \bx - \by|}\rd
    \bx\, \rd \by \;,
\end{eqnarray*}
which shows Eq.~(\ref{eq:3}).  To prove Eq.~(\ref{eq:q1}), we remark that by 
Hilbert's inequality in two dimensions, i.e.,
$|\nabla|\geq h |x|^{-1}$ with $h:=4\pi^2/(\Gamma(1/4))^4$, 
and the diamagnetic inequality \cite{diamagnetic}, we have
  \begin{equation} \label{eq:4}
    \int \rd x|\xi_\mu(x)|^2\int \rd y\frac{|\xi'_\nu(y)|^2} {|\bx-\by|}\leq
    h^{-1}(\xi'_\nu,|-\ri\nabla+\bA_0|\xi'_\nu)\;.
  \end{equation}
  The claimed bound follows now by multiplication with
  $|\lambda_\mu\lambda'_\nu|$ and summation over $\mu$ and $\nu$.  $\hfill\Box$

We note that because of Eq.~(\ref{eq:q1}) and Lemma \ref{Ede-Ungleichung}:
\begin{lemma} \label{sec:definition-problem-1} For $\gamma\in F$, we have
\[
(2h)^{-1}\|\gamma\|_1 \left [ \| \gamma \|_F
+ (\mu+B^{1/2}) \|\gamma \|_1 \right ] \geq Q[|\gamma|]\geq 0\;.
\]
\end{lemma}

In a mean-field picture, relativistic electrons are described by
one-particle density matrices $\gamma$ with certain additional
properties. 
In particular, since electrons are fermions, they obey the Pauli principle
and cannot occupy states in the Dirac sea which is given by the
negative spectral subspace of a one-particle Dirac operator
$D_\mathcal{A}$ with electromagnetic vector potential $\mathcal{A}$.
Mathematically, this is implemented by requiring that $0\leq\gamma\leq
\Lambda_\mathcal{A}^+:=\chi_{(0,\infty)}(D_\mathcal{A})$. As indicated
already above, we will choose $\mathcal{A}:=A$, a choice known as the
Furry picture. It is then useful to introduce several sets of 
one-particle density matrices $\gamma$ for the subsequent discussion.
\begin{definition}
  We define the following sets of one-particle density matrices for
  given (maximal) particle number $q\in \rz_+$ 
  \begin{eqnarray}
    \label{set:1pdm}
    \Sd &:=& \{ \gamma\in F\ |\ 0 \leq \gamma \leq \la^+\}\;,\\
    \label{set:1pdmq}
    \Sd_q &:=& \{ \gamma\in \Sd\ |\ 0\leq \tr(\gamma)\leq q\}\;,\\
    \label{set:1pdmdq}
    \Sd_{\partial q}&:=& \{\gamma\in \Sd\ |\ \tr(\gamma)=q\}\;.
  \end{eqnarray}
\end{definition}
We note that all sets are closed subsets of $F$. Furthermore, the
first two are convex. They are only of technical importance, whereas we are
ultimately interested in describing a system with a fixed number of 
electrons $q$, i.e., in minimizing the energy over the set $\Sd_{\partial q}$,
eventually for the quantized case $q=N \in \nz$.
In physical terms, the projection $1-\Lambda^+_A$ can be
interpreted as the one-particle density matrix of the Dirac sea which
we consider frozen.  The energy of such a system in Hartree-Fock
approximation is given by the functional $\cE:F\to\rz$ as
\begin{equation} \label{def1}
  \cE(\gamma)  = \tr[(\Dd)\gamma] +    \alpha \ Q[\gamma].
\end{equation}
The corresponding relativistic model has been successfully used in atomic and
molecular physics \cite{Sucher1980,Sucher1984,Sucher1987,ReiherWolf2009}.
This projection approach implies that all negative-energy
states and the zero modes are occupied, 
and $N$ additional particles are then added on top of the filled
Dirac sea, see also Ref.~\cite{hausler}.  

We now address the mathematical properties of $\cE(\gamma)$ in a magnetically
confined graphene nanostructure.
\begin{lemma} \label{welldefined}
  The energy functional $\cE$ is well defined and continuous in the
  $\|\cdot\|_F$ norm. Furthermore, $\cE(\gamma) \geq -\mu q$ for
  $\gamma \in \Sd_q$.
\end{lemma}
 {\it Proof.} The first two claims follow from the definition of the norm,
  Lemma \ref{sec:definition-problem-1}.
  The lower bound is immediate since the only negative term is
  $-\mu\tr\gamma$. $\hfill\Box$
\begin{lemma}
  \label{koerziv}
  The energy functional $\cE$ is coercive on $\Sd_q$, i.e.,
  $\cE(\gamma_n)\to\infty$ if $\gamma_n\in\Sd_q$ and $\|\gamma_n\|_F
  \to\infty$.
\end{lemma}
{\it Proof.}
For $\psi$ in the domain of $D_A$, which equals the one of $\di$,
we have because of the relative
compactness of the perturbing term $P$ that
\[
  \|D_A\psi\| \geq \|(\di)\psi\|-\|P\psi\|
  \geq (1-\epsilon) \|(\di)\psi\| - M \|\psi\|
\]
for an arbitrarily small positive $\epsilon$ and some $M\in\rz$. Thus,
squaring the inequality and taking operator square roots, we get
$|D_A|\geq c_1 |\di| - c_2$ for some positive constants $c_1$ and $c_2$.
Thus for $\gamma\in \Sd$,
\[
    \tr[(\Dd)\gamma] = \tr((|D_A|-\mu)\gamma) \geq
    c_1\|\gamma\|_F - (c_2+\mu)q\;,
\]
which implies the coercivity, since $Q$ is non-negative on $\Sd$. $\hfill\Box$

In order to fulfill the relative compactness requirement (\ref{eq:P}), it is
in fact sufficient to show relative compactness with respect to the free
Weyl operator (without the magnetic field).  Since the following discussion
does not rely on this result, it has been relegated to the Appendix.

\subsection{Minimization of the Energy\label{s1}}

We now follow Barbaroux \textit{et al.} \cite{Barbarouxetal2005} and 
wish to show the existence of a minimizer for
 the Hartree-Fock energy functional (\ref{def1}). Here we
will consider only particle numbers $q$ which are so small that always
\begin{equation} \label{eq:gleichheit}
  \inf\cE\left(\Sd_q\right)=\inf\cE\left(\Sd_{\partial q}\right)\;,
\end{equation}
i.e., the magnetic dot is not yet saturated with electrons.
Our proof strategy is standard: First we show that it is enough
to minimize over density matrices of finite rank. Then we show that
there is a projection with the same particle number yielding a lower
energy unless it is itself a projection. Moreover, the particle number
is automatically quantized so that it is enough to require $q\in \nz$.
Equation (\ref{eq:gleichheit}) thus corresponds to the case $N\le N_c$
discussed in the Introduction.
Finally, the existence of a minimizer follows by a compactness argument. 
Let us now go through the steps of the proof.

\subsubsection{Reduction to Density Matrices with Finite Spectrum.} 

\begin{lemma}\label{red}
  Assume $\gamma \in
  \Sd_{\partial q}$. Then there exists a sequence of finite rank
  density matrices $\gamma_K\in \Sd_{\partial q}$ such that
  $\|\gamma_K-\gamma\|_F\to0$ as $K\to\infty$.
\end{lemma}
{\it Proof.} Let $\xi_k$, $k\in \nz$ be a complete set of 
eigenfunctions of $\gamma$.
If all eigenvalues are $0$ or $1$, the claim is immediate since
then $\gamma$ is of finite rank itself, and  $\gamma$ is trace
class. Thus we can assume that there is an eigenvalue
$\lambda_n\in(0,1)$.  Now set $\epsilon_K:= q- \sum_{k=1}^K \lambda_k$. 
Then $\epsilon_K$ is a non-negative monotone decreasing sequence tending to zero.
Define $\gamma_K := \sum_{k=1}^K \lambda_k |\xi_k\rangle\langle\xi_k|
+ \epsilon_K|\xi_n\rangle\langle\xi_n|$.  We now assume $n\leq K$ and $K$ so big that
$\lambda_n+\epsilon_K < 1$. Obviously, $0\leq\gamma_K\in\Sd_{\partial q}$ 
and each $\gamma_K$ has finite rank.  We now show that $\gamma_K\to\gamma$ 
in the $F$-norm as $K\to\infty$. We have
\[
 \gamma-\gamma_K = \sum_{k=K+1}^\infty \lambda_k
 |\xi_k\rangle\langle\xi_k| - \epsilon_K|\xi_n\rangle\langle\xi_n| \;.
\]
Thus we obtain
\[
 \|\gamma-\gamma_K\|_F\leq \sum_{k=K+1}^\infty \lambda_k
 \tr(|\pdolch-\mu| |\xi_k\rangle\langle\xi_k|) + \epsilon_K (\xi_n,|\pdolch-\mu|\xi_n)\;.
\]
The first term tends to zero since
$|\pdolch-\mu|^{1/2}\gamma|\pdolch-\mu|^{1/2}\in\gS_1(\gH_A)$, and the
second tends to zero since $\epsilon_K\to0$. $\hfill\Box$

The following is an immediate consequence of the continuity of
$\cE$ in the $F$-norm and the preceding density result:
\begin{corollary} \label{sec:reduct-unren-dens}
  Assume that $q>0$. Then
\[
 \inf\cE_\mu\left(\Sd_{\partial q}\right) =
  \inf \left\{\cE_\mu(\gamma)|\gamma \in \Sd_{\partial q}, \;
  \mathrm{rank}(\gamma)<\infty \right\}\;.
\]
\end{corollary}

\subsubsection{Reduction to Projection.}

In the following we assume $q\geq0$. 
As customary, $[q]:=\max\{k\in\gz| k\leq q\}$ 
denotes the integer part of $q$, and we
set $\epsilon_q:=q-[q]$.  Following the lines of Bach \cite{Bach1992},
we get:
\begin{lemma}
  \label{sec:reduct-proj-1}
  Assume $0\leq\gamma\in \Sd_{\partial q}$ with finite rank. Then
  there exists a projection $\Lambda\in\Sd_{\partial [q]}$ and a self-adjoint
  rank one operator $R$ with $\Lambda R=0$ and $\tr R = 
\epsilon_q$ such that \[ \cE(\Lambda+R)\leq\cE(\gamma)\;. \]
  Equality holds only if $\gamma$ is already of that form.
\end{lemma}
{\it Proof.}  Suppose that $\gamma$ is not of that form. 
Then there exist at least two eigenvalues 
$\lambda,\lambda'\in(0,1)$ of $\gamma$; we denote the corresponding
normalized eigenvectors by $u$ and $v$.  We set
$\tilde\gamma:=\gamma + \epsilon S$,
where $S:=|u\rangle\langle u|-|v\rangle\langle v|$. Note that
$\tilde\gamma\in S_{\partial q}^A$ as long as
$0\leq\lambda+\epsilon\leq1$ and $0\leq \lambda'-\epsilon\leq1$,
which is the case for $\epsilon$ in a neighborhood of zero. We get
\[
    \cE(\tilde\gamma)-\cE(\gamma)=
    \epsilon[\tr(D_AS)+2\Re Q(\gamma,S)] +\epsilon^2Q(S,S)\;.
\]
By explicit computation and use of the Schwarz inequality, we find
$Q(S,S)<0$, since $S$ is a difference of two
orthogonal rank one projections. Now -- depending on the sign of
the coefficient linear in $\epsilon$ -- we lower the energy by
increasing or decreasing $\epsilon$ from zero, until one of the
constraints $ 0\leq \lambda + \epsilon, \lambda'-\epsilon\leq1$ forbids
any further increase or decrease of $\epsilon$.  This process
leaves all the eigenvalues of $\gamma$ unchanged except for $\lambda$
and $\lambda'$, one of which becomes either $0$ or $1$.
Since there are only finitely many eigenvalues of $\gamma$ strictly
between zero and one, even if they are counted according to their
multiplicity, iterating this process eliminates all eigenvalues that
are strictly between $0$ and $1$, i.e., we have found a density matrix
$\Lambda$ such that $\Lambda^2=\Lambda$. $\hfill\Box$

\subsubsection{Criterion for Maximal Charge}

\begin{lemma}\label{t1}
  Assume that for $\gamma\in \Sd_q$ with $\tr\gamma<q$, the
  operator $\la^+ h^{(\gamma)}_\mathrm{HF}\la^+$ has at least $q$ negative
  eigenvalues. Then
  \[
  \inf\cE\left(\Sd_{\partial q}\right)=  \inf\cE\left(\Sd_q\right)\;.
  \]
  If in addition $0\leq\tilde\gamma$ is a minimizer of $\cE$ in $\Sd_q$, it
  follows that $\tr \tilde\gamma=q$.
\end{lemma}
 {\it Proof.} That the left side bounds the right side from above is obvious. To
  prove the converse inequality, we assume that $0\leq\gamma\in \Sd_q$
  with $\tr \gamma < q$. By Lemma~\ref{sec:reduct-proj-1}, we can
  assume that $\gamma$ is a projection $\Lambda$ plus a rank one operator.
  In particular, the dimension of the range of $\Lambda$ is at most $[\tr\gamma]$.  Since the dimension of the
  discrete spectral subspace $\mathfrak{X}$ of $\la^+
 h^{(\gamma)}_\mathrm{HF}\la^+$
  is larger than $q$, we can find $u\in \mathfrak{X}\cap \Lambda(\gH_A)^\perp$
  with $\|u\|\leq1$ and define $\tilde\gamma:=\gamma+\omega$ with
  $\omega:=|u\rangle\langle u|$. We then get
\[
  \cE(\gamma+\omega)-\cE(\gamma) 
=\tr[(D_A-\mu) \omega]+ 2\Re Q(\gamma,\omega)
  =\langle u, h^{(\gamma)}_\mathrm{HF} u\rangle<0\;.
\]
Therefore this construction yields a density matrix $\tilde\gamma$
with strictly smaller energy and a trace that can be made larger
by $\min\{1,q-\tr\gamma\}$. Iteration of the construction yields the
desired result. This proves both claims.       $\hfill\Box$

\subsubsection{Existence of a Minimizer}

We now wish to show the existence of a minimizer by weak lower
semi-continuity of the functional on a minimizing sequence and weak
compactness. This has been addressed by Lieb and Simon
\cite{LiebSimon} in the context of orbitals in the
non-relativistic setting.  For density matrices, it was 
addressed by Solovej \cite{Solovej1991} in the non-relativistic context, 
and by Barbaroux \textit{et al.} \cite{Barbarouxetal2005} 
for relativistic systems.  
\begin{theorem}\label{existence}
  Assume $0\leq q$ and let $\mu$ be in the intersection of the
resolvent set of $D_A$ and the interval $(0,l_1)$, where $l_1$
  is the first positive eigenvalue of $\pndolch$. Then there exists a
  $\gamma\in\Sd_q$ such that
  \[
   \cE(\gamma)=\inf\cE\left(\Sd_{q}\right) \;.
  \]
  Moreover, $\gamma=\Lambda + |\xi\rangle\langle\xi|$ with
  $\Lambda$ a projection, $\Lambda\xi=0$, and $\|\xi\|<1$.
\end{theorem}
{\it Proof.} Let $\gamma_n$ be a 
minimizing sequence in $\Sd_{q}$, i.e.,  
 $\cE(\gamma_n)$ converges to $\inf\cE\left(\Sd_{q}\right)$. Because of the
coercivity of $\cE$ on $\Sd_q$ (Lemma \ref{koerziv}), the sequence
$\gamma_n$ is bounded in $F$.  Thus, according to Banach and
Alaoglu, $\gamma_n$ -- if necessary by extracting a subsequence --
converges in the weak-$*$ topology, i.e., there exist
$\gamma_\infty\in F$ such that for all compact $K$, we have
\[
\tr(K|\di|^{1/2}\gamma_n|\di|^{1/2})\to \tr(K|\di|^{1/2}
\gamma_\infty|\di|^{1/2}).
\]
Since
\begin{eqnarray*}
&& \langle\psi,\gamma_\infty\psi\rangle = \\ && =
  \tr (|\di|^{-1/2}|\psi\rangle\langle\psi| |\di|^{-1/2}|\di|^{1/2}\gamma_\infty|\di|^{1/2} )\\
 && =\lim_{n\to\infty} \tr 
(|\di|^{-1/2}|\psi\rangle\langle\psi|
 |\di|^{-1/2}|\di|^{1/2}\gamma_n|\di|^{1/2} ) \\ &&= 
\lim_{n\to\infty}\langle\psi,\gamma_n\psi\rangle\geq0 \;,
\end{eqnarray*}
we have $\gamma_\infty\geq 0$. Similarly,  $\gamma_\infty\leq \la^+$.
 Picking  an orthonormal basis $e_1,e_2,\ldots$, Fatou's lemma 
 gives -- possibly under extraction of yet another subsequence --
\[
  q\geq\lim_{n\to\infty}\tr\gamma_n \geq \sum_\nu \liminf_{n\to\infty}
  \tr(|e_\nu\rangle\langle e_\nu| \gamma_n) = 
\tr\gamma_\infty \;.
 \]
  Thus the trace may only decrease. Joining these results shows that
  $\Sd_q$ is weakly-$*$ closed, i.e., $\gamma_\infty\in \Sd_q$.
  We now show lower semi-continuity in the weak-$*$ topology.  
  Concerning the one-particle part, we set
  $\Lambda_0^-:=\chi_{(-\infty,0]}(\pndolch)$ and compute, 
for some $\vartheta>0$
and smaller than the first positive  eigenvalue of $D_A$,
\begin{equation} \label{eq:proj-diff} 
\Lambda_{0}^-\Lambda_{A}^+ =   \int_\rz\frac{\rd\eta}{2\pi}
  \Lambda_{0}^-(\pndolch-\vartheta+P+\ri\eta)^{-1}
P(\pndolch-\vartheta+\ri\eta)^{-1}
  \;,
  \end{equation}
  where we used \cite[Lemma~5.6]{Kato1966}
 $\mathrm{sgn} H =
  \pi^{-1} \int_{-\infty}^{\infty} (\ri\eta+H)^{-1}\rd\eta$,
meant as the Cauchy principal value
\[
  \mathrm{sgn}(H) = \frac{1}{\pi} \lim_{r\rightarrow\infty}
  \int_{-r}^{+r} (\ri\eta+H)^{-1} \rd\eta
\]
in the strong topology. Equation (\ref{eq:proj-diff}) shows that the
product of these two projections is compact, expressing that the
orthogonality of the positive and negative spectral subspaces is not
perturbed too much by $P$, see Eq.~(\ref{eq:P}). 
Moreover, it is easy to see that 
$K_A:=|\di|^{1/2}\Lambda_{0}^-\Lambda_{A}^+ |\di|^{-1/2}$ is also compact. 
We are now in a position to show the lower semi-continuity of the
one-particle part:
\begin{eqnarray} \label{eq:einteilchen-halb-stet}
  &&\liminf_{n\to\infty}\tr[(\Dd)\gamma_n] = \\
  \nonumber && \liminf_{n\to\infty}\tr\left[(\di)\gamma_n
  +|\di|^{-\frac12}P|\di|^{-\frac12
  }|\di|^{\frac12}\gamma_n|\di|^{\frac12} \right] \\
  \nonumber && = \liminf_{n\to\infty}\tr\left(|\pndolch-\mu|^{1/2}
  \gamma_n|\pndolch-\mu|^{1/2} \right)\\
  \nonumber &&\quad -2\tr \left(K_A^*K_A|\di|^{1/2}\gamma_\infty|\di|^{1/2} 
\right)+\tr(P\gamma_\infty)\;.
  \end{eqnarray}
Now, using Fatou's lemma and picking an 
orthonormal basis $e_1,e_2,\ldots$, we have for the first summand
\begin{eqnarray*}
  & &\liminf_{n\to\infty}\tr(|\di|^{1/2}\gamma_n|\di|^{1/2})\\ 
   && =  \liminf_{n\to\infty}\sum_\nu 
\tr \left[
|e_\nu\rangle\langle e_\nu||\di|^{1/2}\gamma_n |\di|^{1/2} \right]\\  &&
    \geq  \sum_\nu 
\liminf_{n\to\infty}\tr\left[|e_\nu\rangle\langle e_\nu||\di|^{1/2}\gamma_n 
|\di|^{1/2}\right] = \|\gamma_\infty\|_F\;.
  \end{eqnarray*}
  Inserting this into the last line of
  Eq.~(\ref{eq:einteilchen-halb-stet}) and undoing the first steps again
  gives the desired bound $\liminf_{n\to\infty}\tr[(\Dd)\gamma_n] \ge
  \tr[(\Dd)\gamma_\infty]$.
  It remains to show the lower semi-continuity of the interaction
  part which is quadratic in the density matrix. Although it is
  quadratic and positive on $\Sd_q$, it is not a positive quadratic form
  on a vector space, i.e., its lower semi-continuity does not follow
  immediately from the Schwarz inequality. However, we can proceed as
  follows. First, we note that there is some constant $C$ such that
$C\geq\|\gamma_n\|_F=\sum_\nu (\xinn,|\di|\xinn)$, where the $\xinn$ are
  eigenfunctions of $\gamma_n$ with $\|\xinn\|\leq 1$. Because of
  Corallary \ref{sec:reduct-unren-dens} and  Lemma \ref{sec:reduct-proj-1}, we
  can assume that the sum contains at most $[q]+1$ summands. Thus, by
  Lemma \ref{p-pa}, also the standard Sobolev norm of $\chi\xinn$ is
  bounded uniformly in $n$ (and $\nu$) for any $C_0^\infty$ function $\chi$.
  Thus, possibly by extracting another subsequence, we can assume that
  the $\xinn\chi$ converge weakly in the $H^{1/2}$-norm to
  $\xi_1,\ldots,\xi_{[q]+1}$. Thus, $\xinn(x)\to \xi_n(x)$ almost
  everywhere pointwise, see, e.g., Ref.~\cite[Theorem~16.1]{LionsMagenes1972}. 
The pointwise convergence allows us to use Fatou's lemma to show that 
\[
\liminf_{n\to\infty} Q[\gamma_n]\geq
  Q[|\xi_1\rangle\langle\xi_1|+\ldots+|\xi_{[q]+1}
\rangle\langle\xi_{[q]+1}|]\;.
\]
We note that the ways of taking the limits -- pointwise and in
the weak-$*$ sense -- agree, i.e., the pointwise limit of
$\sum_\nu\xinn(x)\overline{\xinn(y)}$ is an integral kernel of
$\gamma_\infty$ \cite{Barbarouxetal2005}. $\hfill\Box$

The critical particle number $N_c$ for which
the minimizer $\gamma$ on $\Sd_N$ has trace $N$ is certainly
positive. Unfortunately, for an arbitrary perturbation $P$,  
we have not been able to obtain effective bounds neither from 
below nor from above for $N_c$.  
In the next section, we therefore consider a 
specific choice for the perturbation $P$, first proposed in
 Ref.~\cite{ademarti},
and discuss our results for the numerical solution of the
self-consistent Hartree-Fock equations.  

\section{Numerical results}
\label{sec4}

In this section, we consider a circularly symmetric magnetic dot in graphene, 
defined by a homogeneous background field
$B$ with vector potential $\bA_0 (\bx)=(B/2)(-x_2,x_1)$ and a
purely magnetic perturbation $P= \boldsymbol\sigma\cdot\bA_1$ with
\begin{equation}\label{specificdot}
\bA_1(\bx) = -(Br/2) [\Theta(R-r)+(R/r)^2 \Theta(r-R)] \left(
\begin{array}{c} -\sin\phi  \\ \cos\phi \end{array}\right)\;,
\end{equation}
where $x_1=r\cos\phi$, $x_2=r\sin\phi$,
and $\Theta$ is the Heaviside function.
This choice implies that the magnetic field vanishes inside a disc of 
radius $R$ around the origin, while outside this disc the field is constant
and given by $B$.
On a semiclassical level, one can expect bound states inside the disc,
built by combining the plane-wave solutions inside the disc with the
cyclotron orbit solutions outside the disc. 
The existence of bound states is shown below under a fully quantum mechanical
approach. Note that single-particle
states centered far away from this disc (``magnetic dot'')  represent 
spatially localized Landau cyclotron orbits. 
The dimensionless ``missing flux'' $\delta:= R^2/2l_B^2$, where 
$l_B: = (c/|eB|)^{1/2}$ is the magnetic length, 
then controls the number of bound positive-energy single-particle
states energetically below the first bulk Landau level.
In what follows, all energies are given in units of the
 first Landau level energy, $\sqrt{2}\hbar v/l_B$.

We solve the interacting problem ($\alpha>0$) numerically within the 
self-consistent Hartree-Fock approximation discussed in Sec.~\ref{sec3},
with density matrices $\gamma$ with $\tr (\gamma)=N$ 
constrained to the Hilbert space spanned by positive 
single-particle energies.
The familiar $\delta=0$ Landau level energies \cite{review2},
$\varepsilon_{s} := \sigma \sqrt{n+(j+1/2)\Theta(j)}$,
are expressed in terms of 
the set $s:=(j,n,\sigma)$ of quantum numbers
 $j\in \gz+1/2$ (angular momentum), $n\in \nz_0$
(radial index), and the conduction/valence
band index $\sigma=\pm$.  (For $n=0$ and $j<0$,  only $\sigma=-$ is allowed 
and spans the zero-energy level $\varepsilon=0$.)  
The corresponding single-particle eigenspinor $|s\rangle$ 
has the spatial representation \cite{ademarti}
\begin{equation}\label{njpsi}
\Psi_{s=(jn\sigma)} (\xi,\phi):= \langle \bx|s\rangle
= \frac{e^{\ri j\phi}}{\sqrt{2\pi}}
\left( \begin{array}{c} e^{-\ri \phi/2} \psi^+_{jn}(\xi)\\
\ri\sigma e^{\ri \phi/2} \psi^-_{jn}(\xi) \end{array} \right)\;,
\end{equation}
where $\xi:= r^2/2l_B^2$ is a dimensionless radial coordinate, 
and we have $\langle s|s'\rangle=\delta_{ss'}$ with
$\int_0^\infty d\xi [(\psi^+_{nj})^2+(\psi^-_{nj})^2]=1.$
Using the generalized Laguerre polynomials $L_n^k$, the 
Landau states (\ref{njpsi}) contain the real-valued functions
\begin{eqnarray}\label{waveexp}
\psi^+_{nj}(\xi) &=& A_{jn}^+ \
 \xi^{\frac12|j-\frac12|} \ e^{-\xi/2} \ L_{n-\Theta(-j)}^{
|j-\frac12|} (\xi) \;, \\ \nonumber 
\psi^-_{nj}(\xi) &=& A_{jn}^- \ \xi^{\frac12|j+\frac12|} \ e^{-\xi/2} \
 L_{n}^{|j+\frac12|} (\xi)\;, 
\end{eqnarray}
with normalization factors 
\[
A_{jn}^+ = \sqrt{\frac{ (n-\Theta(-j))!}{2 (n-\Theta(-j)+|j-\frac12|)!}},\quad
A_{jn}^- = {\rm sgn}(j) \sqrt{\frac{ n!}{ 2 (n+|j+\frac12|)!}}\;.
\]
For $n=0$ and $j<0$, we have $A_{jn}^+=0$
and $A_{jn}^-$ has to be multiplied by $\sqrt{2}$.

\begin{figure}
\begin{center}
\includegraphics[width=0.85\textwidth]{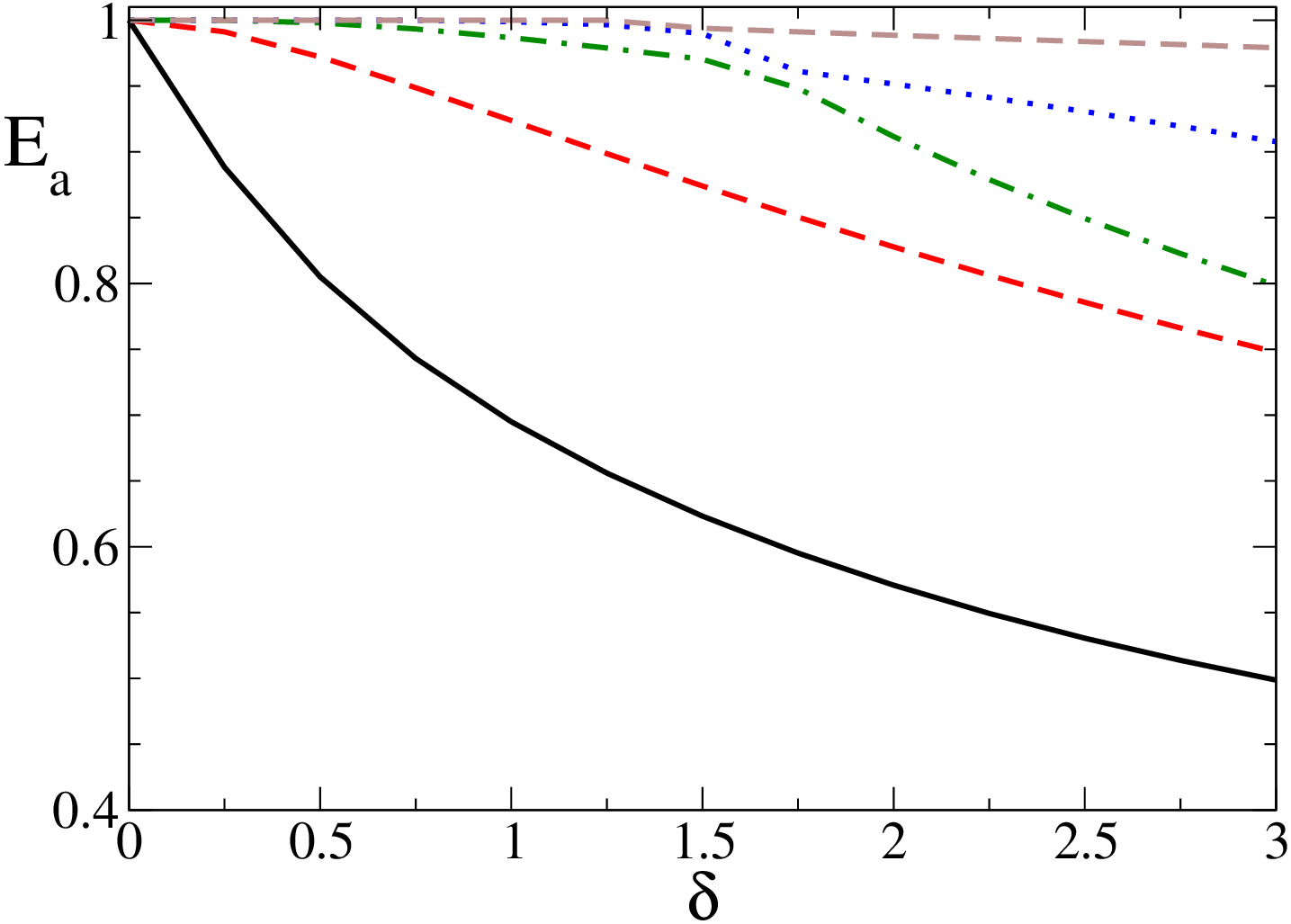}
\end{center}
\caption{\label{f1} Single-particle spectrum of the circular
magnetic dot (\ref{specificdot}) vs dimensionless missing 
flux $\delta$.  The lowest 
five positive-energy solutions $E_a$ are shown; $E_a$ is given in
units of the first Landau level energy.}
\end{figure}

It is then straightforward to express the perturbation $P$ in this basis,
and to diagonalize the full single-particle problem numerically.
Note that $j$ is still a good quantum number, and we label the 
positive energy solutions $|a\rangle$ with energy $E_a$
by $a=(j,k)$ with $k\in \nz$.  
Numerical diagonalization of the single-particle Hamiltonian
yields  the orthogonal matrix $A$ in the expansion
$|a\rangle = \sum_s A_{s,a} |s\rangle$.
Assuming an inert filled Dirac sea, we only keep states
 with $E_a>0$ in what follows.
The zero-energy states are thus included in the filled Dirac
sea, i.e., the chemical potential is assumed to be just above zero.
The energies $E_a$ are shown as a function of the missing flux $\delta$
in Fig.~\ref{f1}. Bound states correspond to 
states with energy $0<E_a<1$ that are localized near the origin.
As shown in Ref.~\cite{sse},  there are only finitely many states
below $\mu$ -- note that $\mu$ as defined after Eq.~(\ref{energie}) is
a constant in $(0,1)$ which we have not yet specified -- whereas there
are infinitely many states with energy between $\mu$ and 1. These states
with energy close to 1 are not localized near the dot and correspond to weakly
perturbed Landau states far away from  the dot. We will now pick $\mu$
such that it implements this intuition.
We have checked that under the choice $\mu=0.99$, all states with $0<E_a<\mu$ 
correspond to bound states localized near the dot while those with 
$\mu <E_a<1$ correspond to states far away from the dot. 
{}From Fig.~\ref{f1}, we can then read off the number of bound 
states within the magnetic dot.  

Next we address the interacting multiparticle problem, where
$N$ electrons are added on top of the filled Dirac sea. 
With the numerically obtained matrix $A$, 
the two-particle interaction matrix elements 
\[
V_{a_1a_2a_3a_4} :=
\sum_{s_1,s_2,s_3,s_4} A_{s_1,a_1} A_{s_2,a_2} 
\tilde V_{s_1s_2s_3s_4} A_{s_3,a_3} A_{s_4,a_4}
\]
follow from the Landau-state matrix elements 
\begin{equation}\label{twopart}
\tilde V_{s_1s_2s_3s_4} = \frac{\alpha}{\sqrt{2}}
\int \frac{d\bx d\bx^\prime}{|\bx -\bx^\prime|} (\Psi^\dagger_{s_4}\cdot
\Psi_{s_1}^{})(\bx) \ (\Psi^\dagger_{s_3}\cdot\Psi^{}_{s_2})(\bx^\prime)\;, 
\end{equation}
where $s_i=(j_i,n_i,\sigma_i)$ and lengths (energies) 
are expressed in units of $l_B$ (the first Landau level energy).
Angular momentum conservation dictates $j_1+j_2=j_3+j_4$, and
numerically we find that only momentum 
exchange processes with $k:=|j_4-j_1|\le 4$
need to be kept, cf.~also Ref.~\cite{hausler}.
A useful but lengthy explicit expression for the matrix
elements (\ref{twopart}) is given in \ref{appb}.

\begin{figure}
\begin{center}
\includegraphics[width=0.85\textwidth]{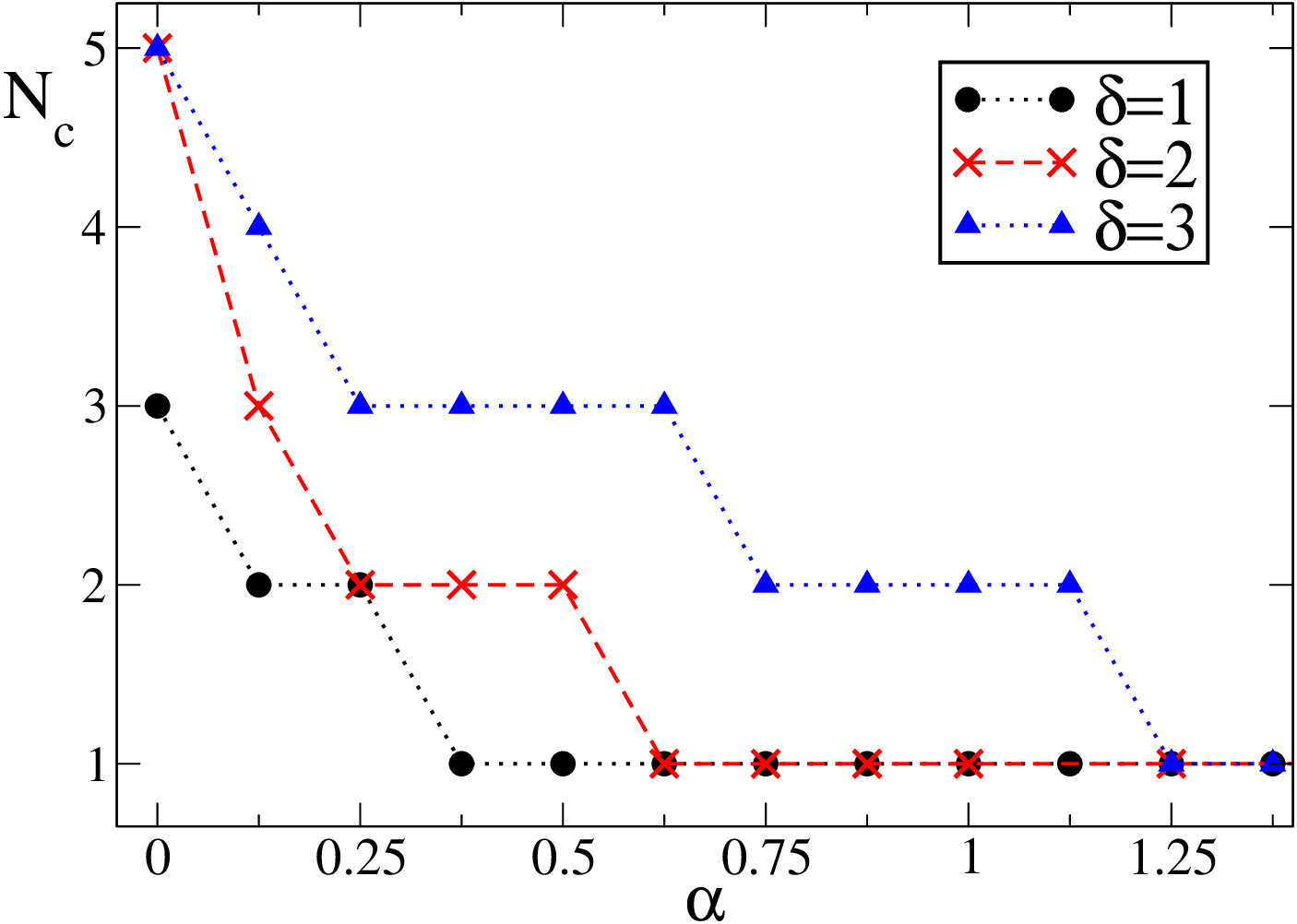}
\end{center}
\caption{\label{f2}  Critical particle number $N_c$ vs interaction
strength $\alpha$ for three different values of the missing 
flux $\delta$.  The symbols represent the computed values, and lines
connecting them are guides to the eye only.  }
\end{figure}

The Hartree-Fock scheme to determine the ground state energy $E_N$ for
$N$ particles with given interaction strength $\alpha$ and
missing flux $\delta$ is then standard \cite{ReiherWolf2009,vignale}. 
Numerical calculations were carried out by restricting the
Hilbert space to $-18<j<2$ and $n=0,1,2$, which spans the relevant
low-energy sector and very accurately describes all stable (i.e., $N\le N_c$)
multiparticle ground-state energies reported below.

The self-consistent numerical solution for the density matrix $\gamma$ 
also allows to read off the number of bound electrons
in the interacting dot. In particular, only
a maximum number $N_c=N_c(\alpha,\delta)$ of electrons can be bound
by the magnetic dot, and for $N>N_c$, we find that $N-N_c$ electrons 
enter bulk Landau states centered far away from the dot (and from each other)
in order to minimize the Coulomb interaction energy. 
The diagonal elements $\gamma_{aa}$ yield the occupation
probability of state $|a\rangle$, and we can thereby directly
infer $N_c(\alpha,\delta)$ from the converged density matrix.
The result is shown in Fig.~\ref{f2} for several values of $\delta$.
Note that $N_c(\delta)$ for $\alpha=0$ follows directly from Fig.~\ref{f1}.
When increasing the repulsive interaction strength $\alpha$, 
electrons tend to be pushed out of the dot, and $N_c$ monotonically
decreases with $\alpha$.  For sufficiently strong interactions, 
$\alpha\ge 1.75$, and moderate values of the missing flux, 
$\delta\le 3$, we find that only a single
electron can be bound by such a dot ($N_c=1$).
Coulomb interactions in graphene nanostructures are thus
very significant for the physically relevant regime $\alpha<2$.

\begin{figure}
\begin{center}
\includegraphics[width=0.85\textwidth]{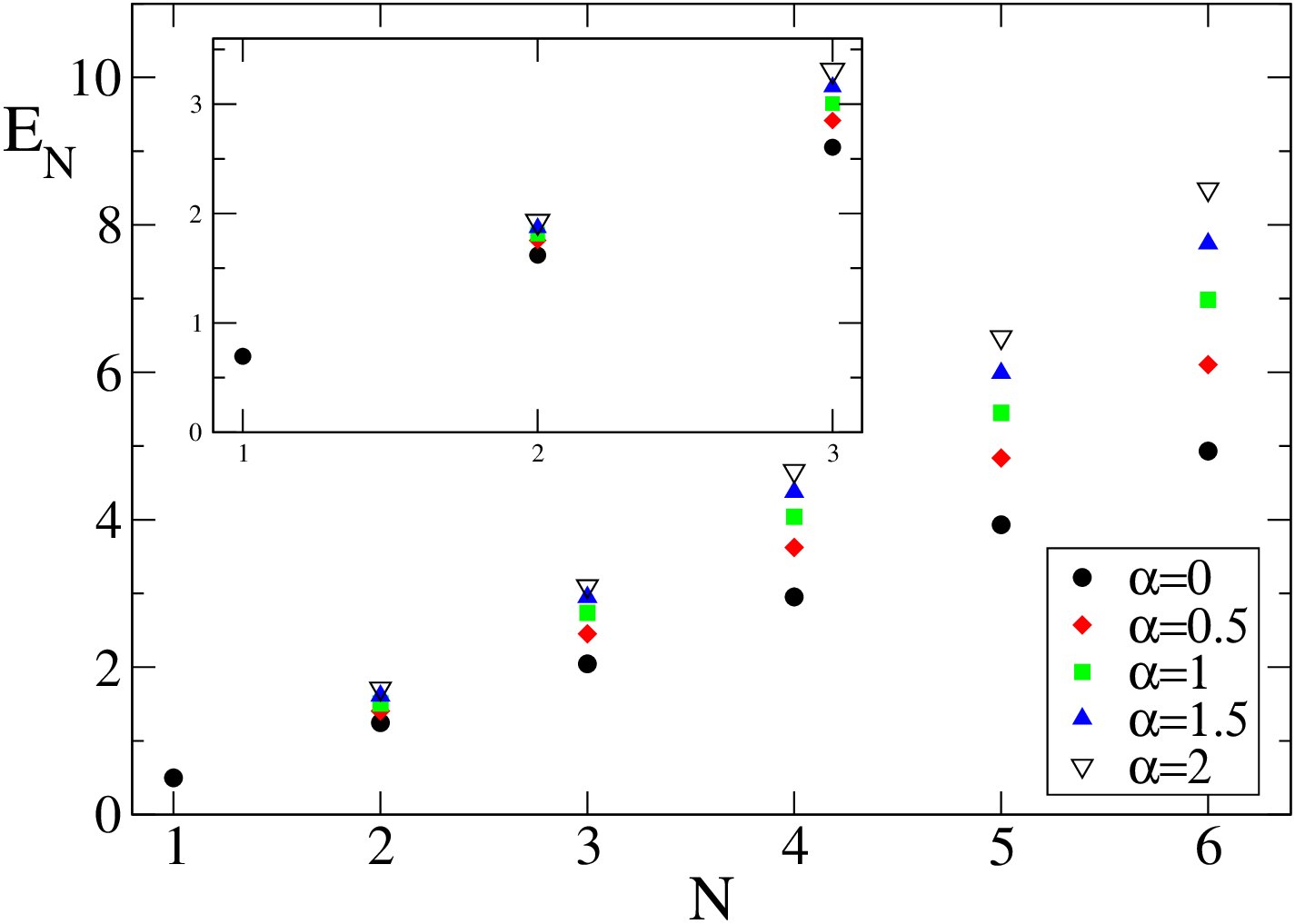}
\end{center}
\caption{\label{f3} Hartree-Fock ground state energy $E_N$
vs $N$ for several $\alpha$ with $\delta=3$.  Inset: Same for
$\delta=1$.}
\end{figure}

Hartree-Fock results for the ground state energy $E_N$ are shown
for several $\delta$ and $\alpha$ in Fig.~\ref{f3}.  For $N>N_c$,
the shown energies still depend on the chosen Hilbert
space dimension, and for bigger basis size, they move to 
smaller values. This is clear on physical grounds, since 
$N-N_c$ electrons will stay as far away as possible
from the dot region and from each other
in order to minimize the Coulomb energy.  
In the limit of infinite basis size, the Coulomb interaction energy can 
be minimized by moving $N-N_c$ electrons to infinity,
such that $E_{N+1}-E_N=1$ for $N\ge N_c$.  Obviously, for the
chosen basis size, the results in Fig.~\ref{f3} do not yet 
match onto this asymptotic form.
However, the shown results for $E_N$ with $N\le N_c$ 
in Fig.~\ref{f3}, describing the stable multiparticle case,
are accurate and do not change when increasing the basis
size.  This behavior gives an additional criterion
to find the number $N_c$ and thereby the 
sought stability condition.

\section{Discussion}\label{sec5}

In this paper, we have studied the interacting
multiparticle problem in graphene quantum dots.  Such quantum dots
can be created in a tunable way by imposing suitable inhomogeneous
magnetic field profiles.  After providing the mathematical 
foundations of Hartree-Fock theory and discussing the conditions
for the existence of a minimizer,
we have given predictions for the maximum number
of bound electrons in a specific example.  
The Hartree-Fock ground state energies
shown here represent upper bounds for the true ground state energy.
An alternative to Hartree-Fock calculations is to employ the
M\"uller functional \cite{muller,siedentop}.
As opposed to the Hartree-Fock functional, the exchange energy is no
longer dominated by $D[\rho_\gamma]$ but instead by the
kinetic energy. In fact, the M\"uller correlation energy shares the
feature of the Dirac exchange energy $\rho_\gamma^{4/3}$ that it
underestimates the correlation energy, resulting in too low energies. 
We expect that the M\"uller functional then implies lower bounds
for the ground-state energy, and we can thereby get both upper and
lower bounds for the exact result.  This work is currently in progress.  

\ack

We acknowledge support by the SFB Transregio 12 of the DFG. 

\appendix

\section{Some useful lemmata}\label{a-r}

The relevance of the following result is that it 
suffices to show relative (form) compactness with respect to the 
free Weyl operator to fulfill the compactness requirements (\ref{eq:P}).

For some $R>1$ (not to be confused with the radius $R$
used in Sec.~\ref{sec4}), we define a smooth cutoff function $\chi_{R}\in
C_0^\infty(\rz^2;[0,1])$ with the property that $\chi_{R}=1$ for
$|\bx|\le R/2$ and $\chi_{R}=0$ for $|\bx|\ge R$; it follows that
$\|\nabla \chi_{R}\|_\infty=\mathcal{O}(1/R)$. We will often use that
for $\varphi\in\core$,
\begin{equation} \label{eq:1}
  [D_A,\chi_R]\varphi=-\ri\boldsymbol\sigma \cdot\nabla\chi_R\varphi\;.
\end{equation}
\begin{lemma}\label{relativ-kompakt}
  Let $A=(A_1,A_2)$ be a magnetic vector potential with $A_j\in
  L_{\mathrm{loc}}^\infty(\rz^2)$ for $j=1,2$.  
Let $T$ be a bounded symmetric (matrix-valued) multiplication
operator such that
\begin{itemize}
\item[a)] $\|T(\bx)\|\to 0$ as $\|\bx\|\to \infty$,
\item[b)] $T |\boldsymbol\sigma\cdot\bp+\ri|^{-1/2}$ is a compact
  operator.
\end{itemize}
Then, for any $\lambda$ in the resolvent set of $D_A$, the 
operator  $T | D_A+\lambda|^{-1/2}$ is compact.
\end{lemma}
{\it Proof.} Clearly the claim follows if we prove that
 $\chi_R T|D_A+\ri|^{-1/2} $ is compact for all $R>1$, since 
$\chi_R T |D_A+\ri|^{-1/2}\to T |D_A+\ri|^{-1/2}$ in the operator
norm as $R\to\infty$ and the operator 
$|D_A+\ri|^{1/2}|D_A+\lambda|^{-1/2}$ is bounded. We observe that
\begin{eqnarray*}
&&\chi_R T|D_A+\ri|^{-1/2}=\chi_R T|\boldsymbol\sigma\cdot\bp+\ri|^{-1/2}+\\ 
&&+ T|\boldsymbol\sigma\cdot\bp+\ri|^{-1/2}
|\boldsymbol\sigma\cdot\bp+\ri|^{1/2}
\chi_R (|D_A+\ri|^{-1/2}-|\boldsymbol\sigma\cdot\bp+\ri|^{-1/2})\;.
\end{eqnarray*}
Therefore, it suffices to show that $|\boldsymbol\sigma\cdot\bp+\ri|^{1/2}
\chi_R (|D_A+\ri|^{-1/2}-|\boldsymbol\sigma\cdot\bp+\ri|^{-1/2})$ is a 
bounded operator. 
In order to do so, we first note that the following resolvent 
identity holds in $L^2(\rz^2:\cz^2)$: 
For $z\in \ri \rz\setminus\{0\}$, write
\[
  \ra=(D_A+z)^{-1},\qquad\rp=(\boldsymbol\sigma\cdot\bp+z)^{-1}\;,
\]
then
\begin{eqnarray} \label{eq:7} &&
\chi_R  (\ra -\rp)\\ \nonumber &&
=\rp(\ri\boldsymbol\sigma \cdot\nabla\chi_R)(\ra -\rp)-\rp\boldsymbol\sigma\cdot A\chi_R\ra \\ \nonumber &&
=:R_0(z)\Gamma(z)\;.
\end{eqnarray}
Before proving the above equation, using the spectral
theorem, we compute
\begin{eqnarray*}
&& | D_A+\ri|^{-1/2}=(D_A^2+1)^{-1/4}=
\frac{1}{\sqrt{2} \pi}\int_0^\infty \frac{\rd t}
{t^{1/4}}\frac{1}{D_A^2+1+t}                     \\
&=&\frac{1}{\sqrt{2}  \pi}\int_0^\infty 
\frac{\rd t}{2\ri t^{1/4}\sqrt{1+t}}\left [R_A(-\ri\sqrt{1+t} )-
R_A(\ri\sqrt{1+t} )\right]\;.
\end{eqnarray*}
An analogous formula holds for $|\boldsymbol\sigma\cdot\bp+\ri|^{-1/2}$ with
$R_A$ replaced by $R_0$. Therefore, using also Eq.~(\ref{eq:7}) we get
\begin{eqnarray*}
&&|\boldsymbol\sigma\cdot\bp+\ri|^{1/2}
\chi_R \left(|D_A+\ri|^{-1/2}-|\boldsymbol\sigma\cdot\bp+\ri|^{-1/2}\right)=\\
&&\frac{-1}{\sqrt{2} \pi}
\sum_{\kappa=-1,1}\kappa \int_0^\infty 
\frac{\rd t}{2\ri t^{1/4}\sqrt{1+t}}
|\boldsymbol\sigma\cdot\bp+\ri|^{1/2}R_0(\kappa\ri\sqrt{1+t} )
\Gamma(\kappa\ri\sqrt{1+t})\;.
\end{eqnarray*}
Noting that $|\boldsymbol\sigma\cdot\bp+\ri|^{1/2}R_0(\kappa\ri\sqrt{1+t} )$
is bounded and that $\|\Gamma(\kappa\ri\sqrt{1+t})\|\le c/\sqrt{1+t}$
for some constant $c$, we conclude that the operator above is bounded.

It remains to show Eq.~(\ref{eq:7}).
For $\phi\in\core$,  using Eq.~(\ref{eq:1}) we find
\begin{eqnarray*}
  && (\ra -\rp)\chi_R(\boldsymbol\sigma\cdot\bp+z)\phi \\
&& =(\ra
  -\rp)[\ri\boldsymbol\sigma
  \cdot\nabla\chi_R+(\boldsymbol\sigma\cdot\bp+z)\chi_R]\phi\\ &
 &= (\ra -\rp)\ri\boldsymbol\sigma
  \cdot\nabla\chi_R\phi-\ra  \boldsymbol\sigma\cdot A  \chi_R\phi\\ &&=[(\ra
  -\rp)\ri\boldsymbol\sigma \cdot\nabla\chi_R\rp \\ &&
-\ra  \boldsymbol\sigma\cdot A \chi_R\rp](\boldsymbol\sigma\cdot\bp+z)\phi\;.
\end{eqnarray*}
Since the range of $(\boldsymbol\sigma\cdot\bp+z)| \core$ is dense in
$L^2(\rz^2:\cz^2)$, we obtain the adjoint of Eq.~(\ref{eq:7}) by a limiting
argument.  $\hfill\Box$

The next result allows us to use weak 
$H^{1/2}$ convergence in the proof of the main theorem.
\begin{lemma}\label{p-pa}
  Assume for the magnetic field and the magnetic vector potential that
  their components are locally bounded, i.e., $A_j,B_j\in
  L_{\mathrm{loc}}^\infty(\rz^2)$ for $j=1,2$. Then, for any $\chi\in
  C_0^\infty(\rz^2;[0,1])$, there exists a constant $c_\chi>0$ such
  that for all $\phi\in \core$, we have
\begin{equation} \label{eq:9}
 (\phi,\chi|\bp|\chi \phi)\le 2^{3/2}(\phi,
|\pdolch-\mu|\phi)+c_\chi\|\phi\|^2\;.
\end{equation}
\end{lemma}
{\it Proof.} A simple application of the
inequality $(a+b)^2\ge (1-\delta)a^2+(1-1/\delta)b^2$
yields, as quadratic forms on $\core$,
\begin{eqnarray*}
  && \chi(\pdolch-\mu)^2\chi\ge (1-\delta)\chi
    [(\bp+\bA)^2+\boldsymbol\sigma\cdot\mathbf{B}]\chi+
    (1-\delta^{-1})\mu^2\chi^2\\
  &&  \ge (1-\delta)^2\chi\bp^2\chi+(1-\delta)(1-\delta^{-1})\chi \bA^2
\chi \\ && \quad +(1-\delta)\chi \boldsymbol\sigma\cdot\mathbf{B}\chi+
(1-\delta^{-1})\mu^2\chi^2\\
   &&  \ge \frac{1}{4}\chi\bp^2\chi-c^2\;,
\end{eqnarray*}
where in the last inequality we set $\delta=1/2$ and
$c^2=\|\chi\bA^2\chi\|_\infty+
\|\chi|\mathbf{B}|\chi\|_\infty/2+\mu^2$. With the help of the above
inequality and Eq.~(\ref{eq:1}), we obtain, for any $\phi\in\core$,
\begin{eqnarray*}
 && \frac{1}{4}(\phi,[\chi|\bp|\chi]^2\,
  \phi)\le\frac{1}{4}\||\bp|\chi\phi\|^2\le
  \|(\pdolch-\mu)\chi\phi\|^2+c^2\|\phi\|^2\\
 && \le 2\|(\pdolch-\mu)\phi\|^2+2\|\boldsymbol\sigma\cdot\nabla\chi
  \phi\|^2+c^2\|\phi\|^2 \\ && \le
  2(\phi,(\pdolch-\mu)^2\phi)+(\phi,[2\|\nabla\chi\|_\infty+c^2]\phi)\,.
\end{eqnarray*}
Using this last estimate and the fact that the square root is operator
monotone, we get Eq.~(\ref{eq:9}) with
$c_\chi=2[2\|\nabla\chi\|_\infty+c^2]^{1/2}$. $\hfill\Box$

\section{Two-particle matrix elements}\label{appb}

In this appendix, we provide the explicit form of the 
two-particle Coulomb matrix 
elements (\ref{twopart}) in terms of the Landau level spinors (\ref{njpsi}).
With the radial functions (\ref{waveexp}) and momentum
exchange $k=|j_4-j_1|$, we have
\begin{eqnarray}\label{intnext}
\tilde V_{ s_1s_2s_3s_4 } &=&\frac{\alpha}{2\pi} \int_0^{\pi}d\phi
\int_0^\infty d\xi \int_0^\infty d\xi' \
\frac{\cos(k\phi)}{\sqrt{\xi+\xi'-2\sqrt{\xi\xi'}\cos\phi}} \\ \nonumber
&\times& [ \psi_{s_4}\cdot \psi_{s_1}](\xi) \
 [ \psi_{s_3}\cdot \psi_{s_2}](\xi')\;,
\end{eqnarray}
where $\psi_1\cdot \psi_2 := \psi_1^+ \psi_2^+ + \sigma_1\sigma_2
\psi_1^- \psi_2^-$.  We now employ the expansion formula 
\[
\frac{1}{\sqrt{\xi+\xi'-2\sqrt{\xi\xi'}\ \cos\phi}}
= \frac{1}{\sqrt{\xi_>}} \sum_{\ell=0}^\infty (\xi_</\xi_>)^{\ell/2} P_\ell
(\cos\phi)
\]
with Legendre functions $P_\ell(\cos\phi)$,
where $\xi_>={\rm max}(\xi,\xi')$ and $\xi_<={\rm min}(\xi,\xi')$.
The angular integration can then be performed by using the relation
\begin{equation}\label{integral}
\int_0^\pi \frac{d\phi}{\pi}\ \cos(k\phi) P_\ell(\cos \phi) =
\frac{(2\ell-1)!!}{ 2^{\ell} \ell !} C_{(\ell+k)/2;\ell} \;.
\end{equation}
The coefficients $C_{m;\ell}$ with $m\in \nz_0$
are the expansion coefficients of a hypergeometric function, 
and $C_{(\ell+k)/2;\ell}\ne 0$ only for even $k+\ell$ and $\ell\ge k$.
In particular, $C_{0;\ell}=1$, while for $0<m\le \ell$,
 we have the product representation
\[
C_{m;\ell} = \prod_{i=1}^m \frac{(i-1/2)(\ell+1-i)}{i(\ell+1/2-i)}\;.
\]
To perform the $\xi,\xi'$ integrations in Eq.~(\ref{intnext}),
we insert the explicit form of $\psi^\pm_{nj}(\xi)$ in Eq.~(\ref{waveexp}),
with the generalized Laguerre polynomials ($n,m\in \nz_0$)
\[
L_n^m(\xi) = \sum_{i=0}^n   \frac{1}{i!} 
\left( \begin{array}{c} n+m\\ n-i\end{array}\right ) (-\xi)^i\;.
\]
After some algebra, with the $\ell$ summation only extending
over $\ell+k\in 2\gz$, we find the lengthy result
\begin{eqnarray}\label{vvfinal}
\tilde V_{s_1s_2s_3s_4} &=& \alpha \sum_{\eta,\eta'=\pm}
\sum_{\ell\ge k} \frac{(2\ell-1)!!}{2^{\ell+1} \ell !} C_{(\ell+k)/2;\ell}  \\
\nonumber &\times&
\left(\delta_{\eta,+}+\delta_{\eta,-}\sigma_1\sigma_4\right)
\left(\delta_{\eta',+}+\delta_{\eta',-}\sigma_2\sigma_3\right) 
A_{s_1}^\eta A_{s_2}^{\eta'} A_{s_3}^{\eta'} A_{s_4}^\eta  \\
\nonumber &\times& 
\sum_{m_1=0}^{\tilde n_1} \sum_{m_2=0}^{\tilde n_2}
\sum_{m_3=0}^{\tilde n_3} \sum_{m_4=0}^{\tilde n_4}
\frac{(-)^{m_1+m_2+m_3+m_4}}{m_1 ! m_2 ! m_3 ! m_4 !} \\
&\times&\nonumber
\left( \begin{array}{c} \tilde n_1+|j_1-\eta/2|\\
\tilde n_1-m_1  \end{array}\right) 
\left( \begin{array}{c}
 \tilde n_2+|j_2-\eta'/2|\\ \tilde n_2-m_2  \end{array}\right)
\\ &\times&\nonumber
\left( \begin{array}{c}
 \tilde n_3+|j_3-\eta'/2|\\ \tilde n_3-m_3  \end{array}\right)
\left( \begin{array}{c}
 \tilde n_4+|j_4-\eta/2|\\ \tilde n_4-m_4  \end{array}\right)
\\ &\times&  \nonumber  I(M+M',M') + (s_1,s_4) \leftrightarrow (s_2,s_3)\;,
\end{eqnarray}
where $\tilde n_{1,4}:= n_{1,4}-\delta_{\eta,+}\Theta(-j_{1,4}), 
\tilde n_{2,3}:= n_{2,3}-\delta_{\eta',+}\Theta(-j_{2,3})$, and
\begin{eqnarray*}
M &:=&
 m_1+m_4+ \frac12\left( |j_1-\eta/2|+|j_4-\eta/2|-\ell \right)\;,\\
M' &:=& 
 m_2+m_3+ \frac12\left( |j_2-\eta'/2|+|j_3-\eta'/2|+\ell \right)\;.
\end{eqnarray*}
Finally, for $n,n'\in \nz_0$, we have 
\[
I(n,n'):= \frac{\sqrt{\pi}}{2} \frac{(2 n +1)!! }{2^n}
\int_0^1 dy \ \frac{y^{n'}}{(1+y)^{n+3/2}}\;.
\]
This allows for the numerical evaluation
of the Coulomb interaction matrix elements, since all summations
in Eq.~(\ref{vvfinal}) converge rapidly. Finally, note that the 
matrix elements obey the symmetry relations
\begin{equation}
\tilde V_{s_1s_2s_3s_4}= \tilde V_{s_2s_1s_4s_3}= \tilde V_{s_4s_3s_2s_1}\;.
\end{equation}

\end{document}